\documentclass{PoS}

\title{Nature of finite temperature and density phase transitions in many-flavor QCD}

\ShortTitle{Nature of finite temperature and density phase transitions in many-flavor QCD}

\author{\speaker{Shinji Ejiri}
         \\
        Graduate School of Science and Technology, Niigata University,
        Niigata 950-2181, Japan\\
        E-mail: \email{ejiri@muse.sc.niigata-u.ac.jp}}

\author{Norikazu Yamada\\
        KEK Theory Center, Institute of Particle and Nuclear Studies, 
        High Energy Accelerator Research Organization (KEK), 
        Tsukuba 305-0801, Japan\\
        School of High Energy Accelerator Science, The Graduate University
        for Advanced Studies (Sokendai), Tsukuba 305-0801, JapanAffiliation\\
        }

\abstract{
We investigate the phase structure of $(2+N_{\rm f})$-flavor QCD, where two light flavors and $N_{\rm f}$ massive flavors exist, to discuss the feasibility of the electroweak baryogenesis in realistic technicolor scenario and to understand properties of finite density QCD. 
Because an appearance of a first order phase transition at finite temperature is a necessary condition for the baryogenesis, it is important to study the nature of finite temperature phase transition.
Applying the reweighting method, the probability distribution function of the plaquette is calculated in the many-flavor QCD.
Through the shape of the distribution function, we determine the critical mass of heavy flavors terminating the first order region, and find it to become larger with $N_{\rm f}$.
We moreover study the critical line at finite density and the first order region is found to become wider as increasing the chemical potential. 
We discuss how the properties of real (2+1)-flavor QCD at finite temperature and density can be extracted from simulations of many-flavor QCD.

}

\FullConference{31st International Symposium on Lattice Field Theory LATTICE 2013\\
		 July 29 -- August 3, 2013\\
		 Mainz, Germany}

\begin{document}

\section{Introduction: QCD critical point and technicolor models}
\label{sec:intoro}

\begin{figure}[tb]
\begin{center}
\includegraphics[width=75mm]{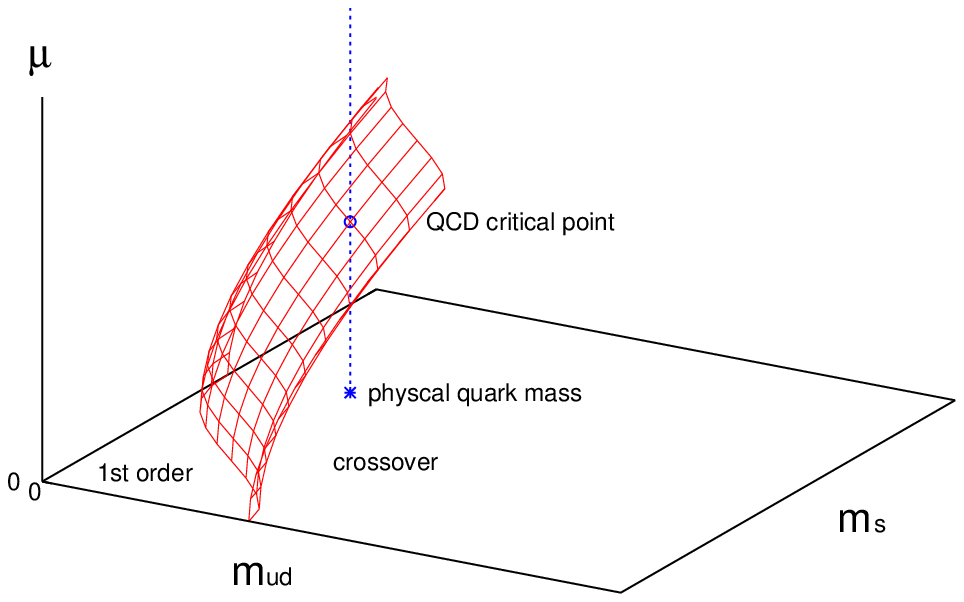}
\includegraphics[width=75mm]{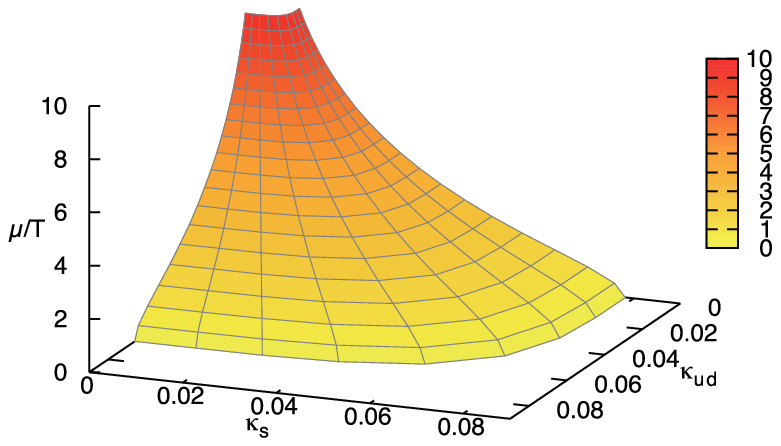}
\vspace{-7mm}
\caption{Left: Expected nature of phase transition in the light quark region of $(m_{ud}, m_s, \mu)$ parameter space.
Right: Critical surface in the heavy quark region obtained in Ref.~\cite{whot2013}. }
\label{fig:crt}
\end{center}
\end{figure}

One of the most interesting topics among current researches of finite temperature QCD is to find the critical point terminating the first order phase transition at finite density. 
From lattice QCD simulations around the physical point of quark masses, the finite temperature phase transition is considered to be crossover at low density, but is expected to turn into a first-order transition when we increase the chemical potential $\mu$ or vary the quark masses.
Since the sign problem becomes severe when we increase $\mu$ in a lattice simulation of QCD, 
it will be important to investigate the location of the critical surface at low density in the quark mass parameter space of $(2+1)$-flavor QCD with dynamical up, down quark masses $m_{ud}$ and strange quark mass $m_s$ 
and to extrapolate the critical surface to the high density region.
The standard expectation for the critical surface as a function of $m_{ud}$ and $m_s$ is illustrated in the left panel of Fig.~\ref{fig:crt}.
The critical surface is indicated by the red curves. 
The left side of the surface near the chiral limit of 3-flavor is the first order region and the right side is crossover. The line of physical quark masses is the blue dashed line.
The point at which this line enters into the first order region is the QCD critical point at finite density. 
A similar critical surface exists also in the heavy quark region, and the location of the critical surface has been computed explicitly at any $\mu$ by Ref.~\cite{whot2013}, which is shown in Fig.~\ref{fig:crt} (right). The colored surface is the critical surface. $\kappa_{ud}$ and $\kappa_s$ are the hopping parameters (inverse masses) of up, down and strange quarks. 
However, recent lattice QCD studies suggest that the critical region of the light quark side at zero density is accessible only when the quark masses are very small   
and thus its determination may be difficult \cite{RBCBi09}.

In this report, we study QCD having two light flavors and many massive flavors \cite{Yamada2012}.
As we will see, the first order transition region becomes wider as the number of massive flavors increases. 
If the critical mass of QCD with many flavors is larger than that of 
$(2+1)$-flavor QCD, the boundary of the first order region can be 
investigated more easily for many-flavor QCD. 
Then, the many-flavor QCD can be a good testing ground for investigating 
$N_{\rm f}$-independent universal properties, such as the critical scaling 
near the tricritical point on the $m_{ud}=0$ axis.
This will provide important information for $(2+1)$-flavor QCD.

Moreover, the study of finite temperature many-flavor QCD is interesting for the construction of the Technicolor (TC) model built of many flavor QCD, i.e.\ vector-like SU(3) gauge theory with many fermions.
In this model, the Higgs sector is replaced by a new strongly interacting gauge theory and its spontaneous chiral symmetry breaking causes electroweak (EW) symmetry breaking.
The EW baryogenesis scenario requires a strong first order phase transition \cite{Appelquist:1995en}. 
It has been known that the phase transition of the standard model is not strong first order \cite{Fodor1999}.
Whereas the nature of the phase transition of many-flavor QCD depends on the number of flavors and masses.

In realistic TC models, two flavors of them are exactly massless and the
mass of other $N_{\rm f}$ flavors must be larger than an appropriate lower
bound otherwise the chiral symmetry breaking produces too many (light pseudo)
Nambu-Goldstone (NG) bosons.
Three of them are absorbed into the longitudinal mode of the weak gauge
bosons, but any other NG bosons have not been observed yet.
On the other hand, the first order transition at small mass terminates
at the critical mass like $(2+1)$-flavor QCD.
Thus, if one requires the first order EW phase transition in TC model, 
it brings in the upper bound on the mass of $N_{\rm f}$ flavors.
This can be a motivation to study $(2+N_{\rm f})$-flavor QCD.

This report is based on Ref.~\cite{Yamada2012}. 
We use a histogram method to identify the order of the phase transitions \cite{Ejiri2008,PRD84,Ejiri2013}. 
The method is explained in the next section.
We then show the result of the endpoint of the first order transition in many-flavor QCD at $\mu=0$ in Sec.~\ref{sec:zeromu}, 
and the chemical potential dependence is discussed in Sec.~\ref{sec:finitemu}. 
Section \ref{sec:summary} is the summary and outlook of this study.

\section{Endpoint of first order phase transitions by a histogram method}
\label{sec:method}

In this study, we consider QCD with two degenerate light quarks of the mass
$m_{\rm l}$ and the chemical potential $\mu$ and $N_{\rm f}$ heavy quarks.
To investigate the nature of phase transitions, we compute the probability distribution function of average plaquette value, 
\begin{eqnarray}
w(P; \beta, m_l, \mu, m_f, \mu_f) 
&=&  \int {\cal D} U {\cal D} \psi {\cal D} \bar{\psi} \
    \delta(P- \hat{P}) \ e^{- S_q - S_g} \nonumber \\
&=& \int {\cal D} U \ \delta(P- \hat{P}) \ 
    e^{6\beta N_{\rm site} \hat{P}}\ (\det M(m_l, \mu))^2 
    \prod_{f=1}^{N_{\rm f}} \det M(m_f, \mu_f)
\label{eq:pdist}
\end{eqnarray}
where $S_g$ and $S_q$ are the actions of gauge and quark fields, 
$M$ is the quark matrix, $N_{\rm site} \equiv N_{\rm s}^3 \times N_t$ 
is the number of site, and $\beta=6/g_0^2$ is the simulation parameter.
$\hat P$ is defined by $\hat P=-S_g/(6N_{\rm site} \beta)$ 
and is $1 \times 1$ Wilson loop for the standard plaquette gauge action.
$\delta (P - \hat{P})$ is the delta function, which constrains the operator $\hat{P}$ to be the value of $P$. 
We moreover define the effective potential,
$V_{\rm eff}(P;\beta,m_f,\mu_f) = -\ln w(P;\beta,m_f,\mu_f)$.

Denoting the potential of 2-flavor at $\mu =0$ by
$V_0(P; \beta)$, 
that of $(2+N_{\rm f})$-flavor 
is written as
\begin{eqnarray}
    V_{\rm eff}(P; \beta, m_f, \mu_f)
=   V_0(P; \beta_0)
  - \ln R(P; \beta, m_f, \mu_f; \beta_0), \ \ 
\label{eq:vefftrans}
\end{eqnarray}
with
\begin{eqnarray}
\ln R(P; \beta, m_f, \mu_f; \beta_0)
&=& 6(\beta - \beta_0)N_{\rm site}P 
 + \ln
     \left\langle
     \displaystyle
     \left( \frac{\det M(m_{\rm l},\mu)}{\det M(m_{\rm l}   ,0)}
     \right)^{2} \prod_{f=1}^{N_{\rm f}}
     \frac{\det M(m_f,  \mu_f)}{\det M(\infty,0)}
     \right\rangle_{\! (P: {\rm fixed})}, \hspace{7mm}
\label{eq:lnr}
\end{eqnarray}
where 
$ \langle \cdots \rangle_{(P: {\rm fixed})} \equiv 
\langle \delta(P- \hat{P}) \cdots \rangle_{\beta_0} /
\langle \delta(P- \hat{P}) \rangle_{\beta_0} $
and $\langle \cdots \rangle_{\beta_0}$ means the ensemble average over
2-flavor configurations generated at $\beta_0$, $m_{\rm l}$ and
vanishing $\mu$.
$\beta_0$ is the simulation point, which may differ 
from $\beta$ in this method.

Restricting the calculation to the heavy quark region,
the second determinant for $N_{\rm f}$ flavors in Eq.~(\ref{eq:lnr}) is approximated
by the leading order of the hopping parameter expansion,
\begin{eqnarray}
\ln \left[ \frac{\det M (\kappa_{\rm h}, \mu_{\rm h})}{\det M (0,0)} \right]  
=  288 N_{\rm site} \kappa_{\rm h}^4 \hat{P} + 12 N_s^3 (2 \kappa_{\rm h})^{N_t} 
\left( \cosh(\mu_{\rm h} /T) \hat{\Omega}_{\rm R} 
+i \sinh(\mu_{\rm h} /T) \hat{\Omega}_{\rm I} \right) + \cdots \ \ 
\label{eq:detmw}
\end{eqnarray}
for the standard Wilson quark action \cite{PRD84}.
$\kappa_{\rm h}$ is the hopping parameter being proportional to $1/m_{\rm h}$, 
$\hat{\Omega}_{\rm R}$ and $\hat{\Omega}_{\rm I}$ are the real and imaginary 
part of the Polyakov loop.
For improved gauge actions such as 
$S_g = -6N_{\rm site} \beta [c_0 {\rm (plaquette)} 
+ c_1 {\rm (rectangle)}]$, 
additional $c_1\times O(\kappa^4)$ terms must be considered, 
where $c_1$ is the improvement coefficient and $c_0=1-8c_1$. 
However, since the improvement term does not affect the physics, 
we will cancel these terms by a shift of the coefficient $c_1$.
The applicability of the hopping parameter expansion will be discussed later.

At a first order transition point, $V_{\rm eff}$ shows 
a double-well shape as a function of $P$, 
and equivalently the curvature of the potential $d^2 V_{\rm eff}/d^2P$ 
takes a negative value in a region of $P$.
To observe this behavior, $\beta$ must be adjusted to be the first 
order transition point.
However, since $\beta$ appears only in the linear term of $P$ in the right 
hand side of Eq.~(\ref{eq:lnr}), 
$d^2V_{\rm eff}/dP^2$ is independent of $\beta$ \cite{Ejiri2008}.
Thus, the fine tuning is not necessary.
Moreover, $d^2V_{\rm eff}/dP^2$ over the wide range of $P$ can be easily
obtained by combining data obtained at different $\beta$ \cite{whot2013}.
We therefore focus on the curvature of
the effective potential to identify the nature of the phase transition.

Denoting $h=2 N_{\rm f} (2\kappa_{\rm h})^{N_t}$ for $N_{\rm f}$ degenerate Wilson quarks with the hopping parameter $\kappa_{\rm h}$, 
we obtain
$\ln R(P;\beta,\kappa_{\rm h},0;\beta_0)
=\ln\bar{R}(P; h,0)$
+(plaquette term) $+ O(\kappa_{\rm h}^{N_t+2})$
for $\mu=\mu_{\rm h}=0$ with
\begin{eqnarray}
\bar{R}(P; h,0)
= \left\langle \exp [6h N_s^3 \hat{\Omega}_{\rm R}] 
\right\rangle_{(P: {\rm fixed}, \beta_0)} .
\label{eq:rew2f}
\end{eqnarray}
$\bar{R}(P; h,0)$ is given by the Polyakov loop and is independent of $\beta_0$.
The plaquette term does not contribute to $d^2V_{\rm eff}/dP^2$
and can be absorbed by shifting  
$\beta \to \beta^{*} \equiv \beta + 48 N_{\rm f} \kappa_{\rm h}^4$ for Wilson quarks. 
Moreover, one can deal with the case with non-degenerate masses by adopting
$h=2 \sum_{f=1}^{N_{\rm f}} (2 \kappa_f)^{N_t}$ for the Wilson quark action or 
$h=(1/4)\sum_{f=1}^{N_{\rm f}} (2m_f)^{-N_t}$ for the staggered quark action.
Thus, the choice of the quark action is not important.
In the following, we discuss the mass dependence of $\bar{R}$ 
through the parameter $h$.

\section{Critical quark mass at zero density for large $N_{\rm f}$}
\label{sec:zeromu}

\begin{figure}[tb]
\begin{center}
\centerline{
\includegraphics[width=75mm]{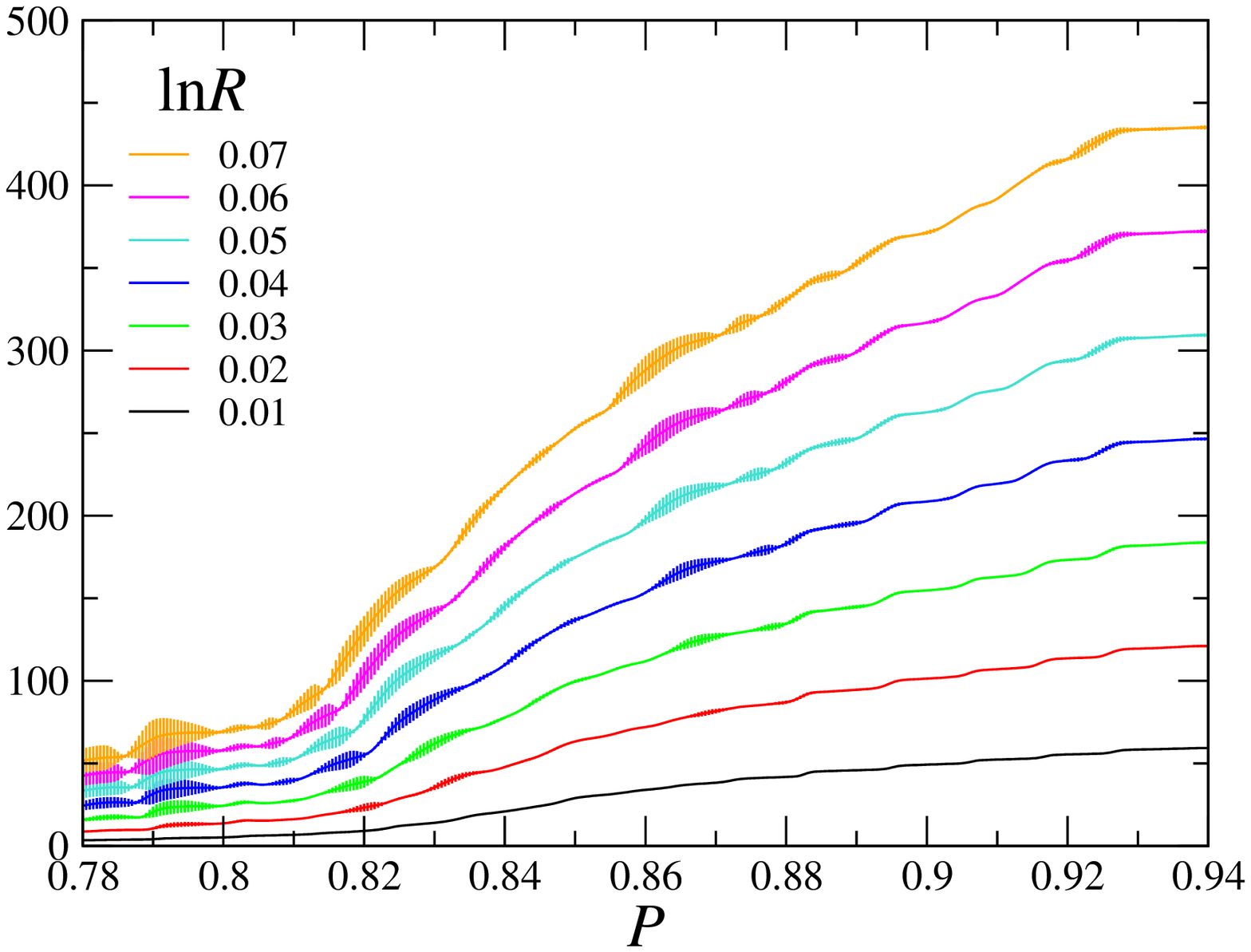}
\includegraphics[width=75mm]{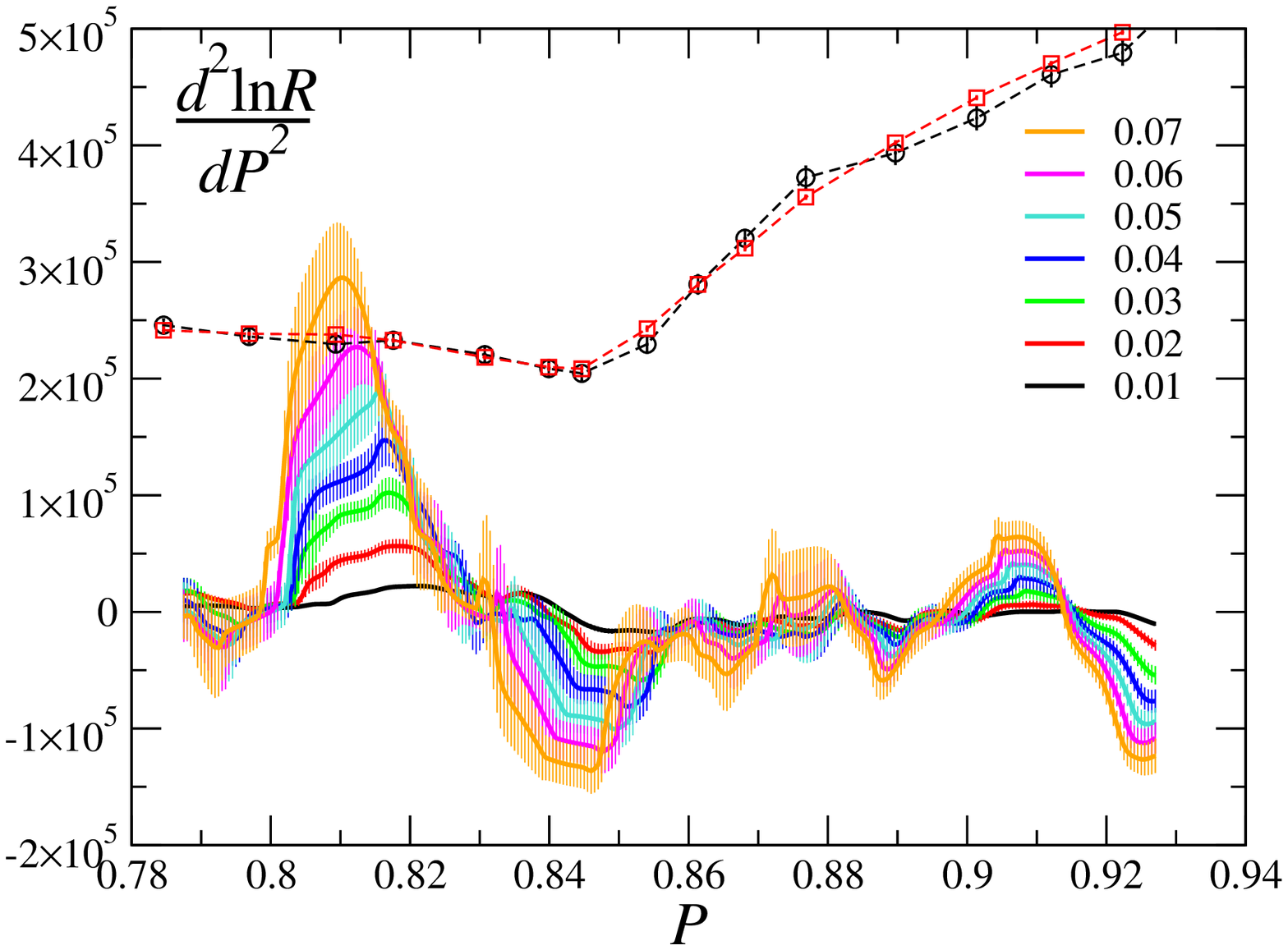}
}
\caption{Left: $\ln \bar{R} (P; h,0)$ as a function of $P$ for $h=0.01$ -- $0.07$.
Right: The colored solid lines are curvature of $\ln \bar{R} (P; h,0)$. 
The dashed lines are $d^2V_0/dP^2(P)$.}
\label{fig:lnr}
\end{center}
\end{figure}

We use the 2-flavor QCD configurations generated with the p4-improved
staggered quark and Symanzik-improved gauge actions \cite{BS05},
thus $\hat P=-S_g/(6N_{\rm site} \beta)$.
The lattice size $N_{\rm site}$ is $16^3 \times 4$. 
The data are obtained at sixteen values of $\beta$ from $\beta=3.52$ to
$4.00$ keeping the bare quark mass to $ma=0.1$.
The number of trajectories is 10000 -- 40000, depending on $\beta$.
The corresponding temperature normalized by the pseudo-critical
temperature is in the range of $T/T_c= 0.76$ to $1.98$, and the
pseudo-critical point is about $\beta=3.65$, where the ratio of
pseudo-scalar and vector meson masses is
$m_{\rm PS}/m_{\rm V} \approx 0.7$.
All configurations are used for the analysis at zero density, while the
finite density analysis is performed every 10 trajectories.
In the calculation of $\bar{R}(P; h,0)$, we use the delta
function approximated by
$\delta(x) \approx 1/(\Delta \sqrt{\pi})$ $\exp[-(x/\Delta)^2]$, 
where $\Delta=0.0025$ is adopted.
Because $\bar{R}(P; h,0)$ is independent of $\beta$, we mix all
data obtained at different $\beta$ as is done in Ref.~\cite{Ejiri2008}.
The results for $\ln \bar R(P;h,0)$ are shown by solid lines in
the left panel of Fig.~\ref{fig:lnr} for $h=0.01$ -- $0.07$. 
A rapid increase is observed around $P \sim 0.82$, and 
the gradient becomes larger as $h$ increases.

The second derivative $d^2 \ln \bar R/dP^2$ is calculated by
fitting $\ln \bar R$ to a quadratic function of $P$ with a range of
$P \pm 0.015$ and repeating with various $P$.
The results are plotted in the right panel of Fig.~\ref{fig:lnr}.
This figure shows that $d^2 (\ln \bar R)/dP^2$ becomes larger with
$h$, and the maximum around $P=0.81$ exceeds $d^2 V_0/dP^2$ for $h > 0.06$.
This indicates that the curvature of the effective potential,
\begin{eqnarray}
\frac{d^2 V_{\rm eff}}{dP^2} = \frac{d^2 V_0}{dP^2} - \frac{d^2 (\ln \bar R)}{dP^2}, 
\end{eqnarray}
vanishes at $h \approx 0.06$ and a region of $P$ where the curvature is negative
appears for large $h$.
We estimated the critical value $h_c$ at which the minimum of
$d^2 V_{\rm eff}/dP^2$ vanishes and obtained $h_c=0.0614(69)$.


We have defined the parameter $h=2 N_{\rm f} (2\kappa_{\rm h})^{N_t}$ for the
Wilson quark.
Then, the critical $\kappa_{hc}$ corresponding $h_c$ decreases as
\begin{eqnarray}
\kappa_{hc}= \frac{1}{2} \left( \frac{h_c}{2N_{\rm f}} \right)^{1/N_t} 
\end{eqnarray}
with $N_{\rm f}$, and 
the truncation error from the higher order terms of the hopping parameter expansion 
in $\kappa_{\rm h}$ becomes smaller as $N_{\rm f}$ increases.
The application range of the hopping parameter expansion was 
examined in quenched QCD simulations with $N_t=4$, by explicitly
measuring the size of the next-to-leading order (NLO) terms of the
expansion~\cite{whot2013}.
They found that the NLO contribution becomes comparable to that in the
leading order at $\kappa_{\rm h} \sim 0.18$.
Hence, this method may be applicable up to around $\kappa_{\rm h}\sim 0.1$.
For instance, in the case of $N_{\rm f}=10$ with $N_t=4$, 
$\kappa_{hc}$ is 0.118.

\section{Returns for understanding the QCD phase transition at finite density}
\label{sec:finitemu}

\begin{figure}[tb]
\begin{center}
\begin{minipage}{7.7cm}
\includegraphics[width=38mm]{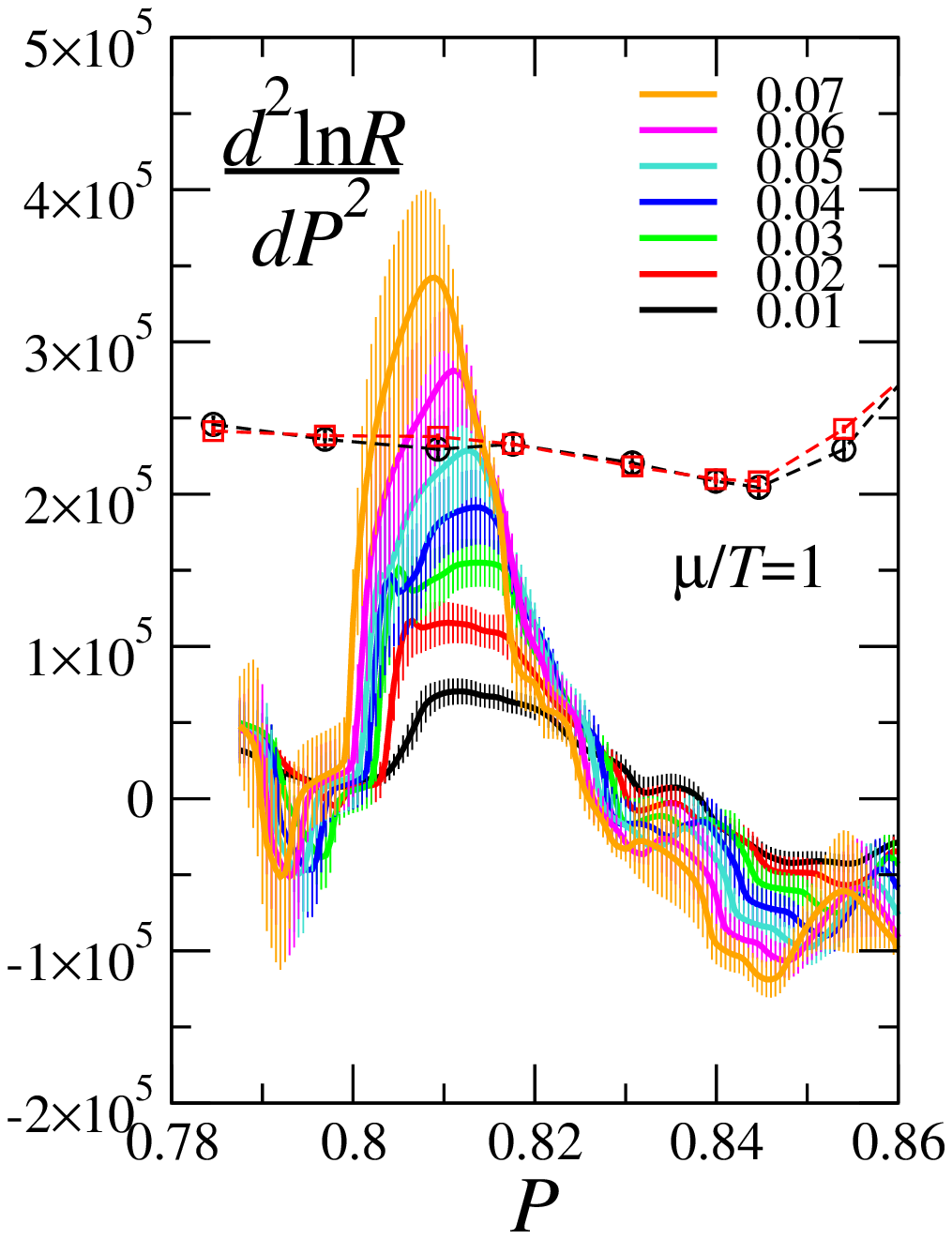}
\includegraphics[width=38mm]{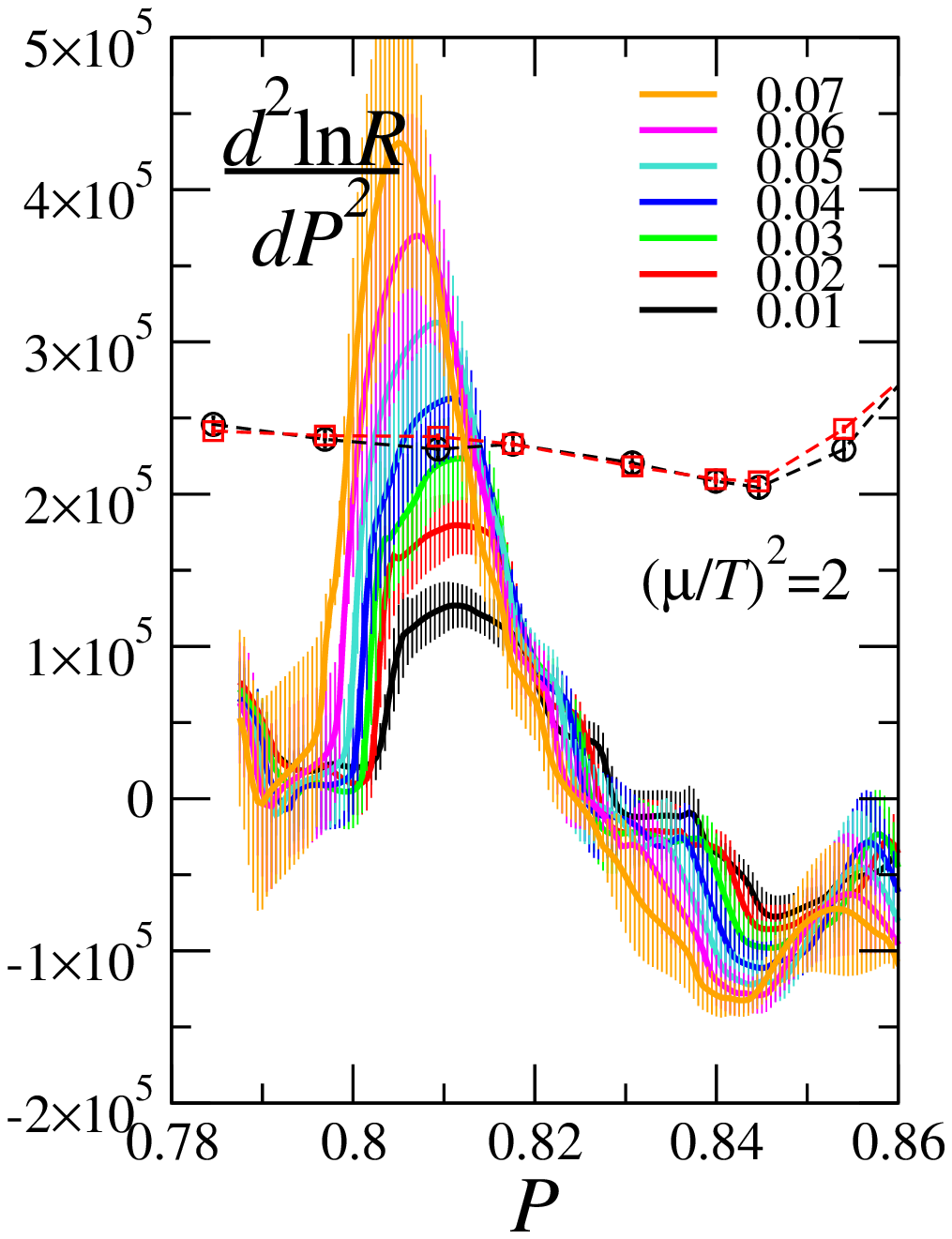}
\hspace{-3mm}
\caption{The curvatures of $\ln \bar{R} (P; h, \mu)$ at 
$\mu /T=1.0$ (left) and $\mu /T=\sqrt{2}$ (right) with $\mu_{\rm h}=0$. 
}
\label{fig:curvmu}
\end{minipage}
\hspace{2mm}
\begin{minipage}{6.8cm}
\includegraphics[width=67mm]{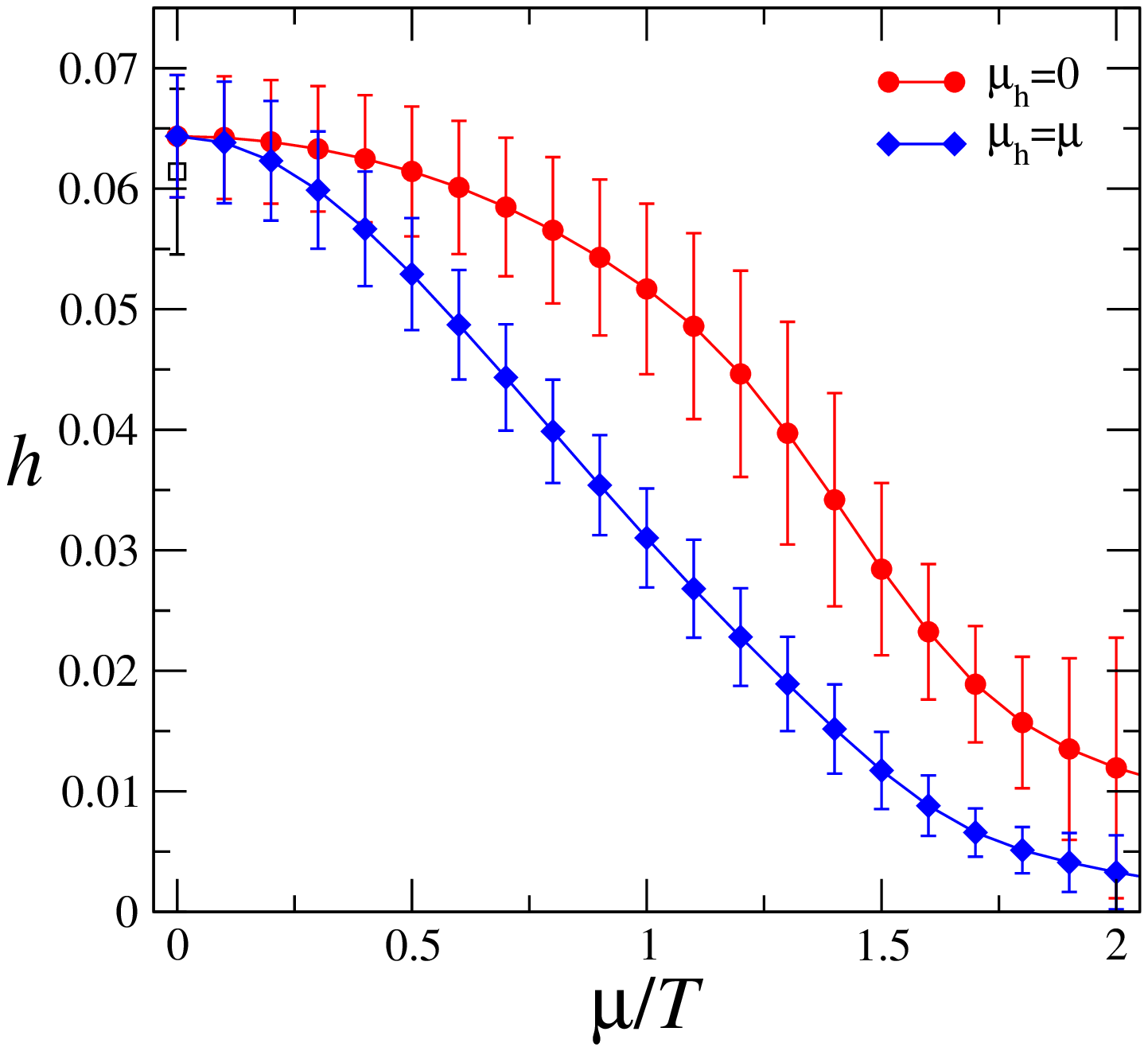}
\vspace{-3mm}
\caption{The critical lines in the $(h, \mu)$ plane for
$\mu_{\rm h}=0$ (circles) and $\mu_{\rm h}=\mu$ (diamonds) \cite{Yamada2012}.
}
\label{fig:crtmu}
\end{minipage}
\end{center}
\end{figure}

We turn on a chemical potential $\mu$ for two light quarks and 
$\mu_{\rm h}$ for $N_{\rm f}$ flavors, and 
discuss the $\mu$ dependence of the critical mass.
As discussed above, we can investigate the critical region more easily for 
large $N_{\rm f}$.
$R(P;h,\mu)$ is then given by Eq.~(\ref{eq:lnr}),
i.e.\ $\langle (\det M(m_{\rm l},\mu)/ \det M(m_{\rm l},0))^2$ $\times$
$(\det M(m_{\rm h},\mu_{\rm h})/ \det M(\infty,0))^{N_{\rm f}} \rangle_{(P: {\rm fixed})}$ for finite $\mu$ and $\mu_{\rm h}$.
The quark determinant is computed using the Taylor expansion of 
$\ln [\det M(m_{\rm l}, \mu)/ \det M(m_{\rm l}, 0)]$ in terms of 
$\mu/T$ up to $O[\mu^6]$ and the Gaussian
approximation in \cite{Ejiri2008} is applied to avoid the sign problem.
This approximation is valid for small $\mu$.
Figures \ref{fig:curvmu} shows the curvatures of $V_0$ and
$\ln \bar R$ at $\mu/T=1$ (left) and $\sqrt{2}$ (right) with $\mu_{\rm h}=0$. 
The maximum values of $d^2 \ln \bar R/ dP^2$ increases with $\mu$.
This means the critical $h$ becomes smaller as $\mu$ increases. 

We compute the critical $h$. 
The circle symbols in Fig.~\ref{fig:crtmu} indicate the critical value of $h$ 
as a function of $\mu$ for $\mu_{\rm h}=0$, and the diamond symbols are those 
for $\mu_{\rm h}=\mu$.
In the region above this critical line, the effective potential has the negative
curvature region, indicating the transition is of first order.
It is clear that the first order region becomes wider as $\mu$ increases.
If the same behavior is observed in $(2+1)$-flavor QCD, 
this gives the strong evidence for the existence of the critical point at
finite density in the real world.

Although this analysis is valid only for large $N_{\rm f}$, it gives a frame of
reference for the study of critical mass at finite $\mu$.
Notice that $\ln \bar{R}(P;h,\mu)$ is given by the sum of
$\ln \bar R(P;0,\mu)$ and $\ln \bar R(P;h,0)$ approximately and that
the behavior of $\ln \bar{R}(P;h,0)$ in Fig.~\ref{fig:lnr} is very similar 
to that of $\ln R(P;0,\mu)$ in Fig.~5 and 7 of Ref.~\cite{Ejiri2008}.
$\ln R(P;0,\mu)$ is estimated from the quark number susceptibility
at small $\mu$ and $\ln \bar{R}(P;h,0)$ is obtained from the Polyakov
loop at small $\kappa_{\rm h}$.
Both the quark number susceptibility and the Polyakov loop rapidly
increase at the same value of $P$ near the transition point, which
enhances the curvature of $\ln R$.
Therefore, the critical $h$ decreases with $\mu$ or 
the critical $\mu$ decreases with $h$.
The similar argument is applicable for $(2+1)$-flavor QCD. 

Moreover, an interesting application is to study 
universal scaling behavior near the tricritical point.
In the case that the chiral phase transition in the two flavor massless 
limit is of second order, the boundary of the first order
transition region $m_{\rm l}^c (m_{\rm h})$ is expected to behave as 
$m_{\rm l}^c \sim |m_{\rm E} -m_h|^{5/2}$
in the vicinity of the tricritical point, 
$(m_{\rm l}, m_{\rm h}, \mu)=(0, m_{\rm E}, 0)$,
from the mean field analysis.
This power behavior is universal for any $N_{\rm f}$. 
The density dependence is important as well, which is expected to be
$m_{\rm l}^c \sim |\mu|^5$  \cite{ejirilat08}.
The study of the light quark mass dependence of the critical mass has been 
started and the preliminary results show that the dependence seems to be small \cite{Yamada2013}.

\section{Summary and outlook}
\label{sec:summary}

We studied the phase structure of (2+$N_{\rm f}$)-flavor QCD
to explore the realizability of the EW baryogenesis in technicolor
scenario and to understand properties of the finite density QCD.
Fixing the mass of two light quarks,
we determined the critical mass of the other $N_{\rm f}$ quarks  
separating the first order and crossover regions.
The critical mass is found to become larger with $N_{\rm f}$.
Furthermore, the chemical potential dependence of the critical mass
is investigated for large $N_{\rm f}$, and the critical mass is found to
increase with $\mu$.
If $(2+1)$-flavor QCD has the same property, 
this gives the strong evidence for the existence of the critical point at
finite density in the real world.
Starting from large $N_{\rm f}$, the systematic study of properties of 
QCD phase transition would be possible.

The next step we must investigate is the light quark mass dependence 
of the critical heavy quark mass \cite{Yamada2013}.
The nature of the chiral phase transition in the 2-flavor massless 
limit is still open question \cite{Aoki2012}. 
In the case that the transition of 2-flavor QCD is of second order and the tricritical point exists, 
the universal scaling behavior near the tricritical point is expected, 
which is independent of $N_{\rm f}$, as discussed above. 
On the other hand, if the transition is of first order, the critical 
$\kappa_{\rm h}$ vanishes before going to the 2-flavor massless limit.
Thus, our approach may give important information to solve the long-standing problem of the chiral phase transition of 2-flavor QCD.

\paragraph{Acknowledgments}
This work is in part supported by Grants-in-Aid of the Japanese Ministry
of Education, Culture, Sports, Science and Technology
(No.\
22740183, 
23540295 
).

\end{document}